\documentstyle[pra,aps,multicol]{revtex}

\newcommand{\be}{\begin{equation}}
\newcommand{\ee}{\end{equation}}
\newcommand{\br}{\begin{eqnarray}}
\newcommand{\er}{\end{eqnarray}}

\newcommand{\bd}{\begin{displaymath}}
\newcommand{\ed}{\end{displaymath}}

\newcommand{\bfig}{\begin{figure}}
\newcommand{\efig}{\end{figure}}

\def\3cdot{\cdot \cdot \cdot}

\def\om0{\omega _0}
\def\Om0{\Omega _0}

\def\text#1{{\rm{#1}}}

\def\->{\rightarrow}
\def\=>{\Rightarrow}
\def\-->{\longrightarrow}
\def\==>{\Longrightarrow}

\def\pr{^\prime}
\def\pr2{^{\prime\prime}}

\def\bfig{\begin{figure}}
\def\efig{\end{figure}}

\begin{document}
\draft
\title{Optical bistability in sideband output modes induced by squeezed vacuum}
\author{L. P. Maia\thanks{%
E-mail: lpmaia@ifsc.sc.usp.br}, G. A. Prataviera\thanks{
E-mail:gap@df.ufscar.br} and S. S. Mizrahi\thanks{
E-mail:salomon@df.ufscar.br}}
\address{$^{\ast }$Instituto de F\'{i}sica de Sao Carlos, USP, S\~{a}o Carlos,\\
13560-970, SP, Brazil.\\
$^{\dagger \,\ddagger }$Departamento de F\'{i}sica, CCET, Universidade\\
Federal de S\~{a}o Carlos,\\
Rodovia Washington Luiz Km 235, S\~ao Carlos, 13565-905, SP, Brazil.\\
}
\date{\today}
\maketitle

\begin{abstract}
We consider $N$ two-level atoms in a ring cavity interacting with a
broadband squeezed vacuum centered at frequency $\omega _{s}$ and an input
monochromatic driving field at frequency $\omega $. We show that, besides
the central mode (at $\omega $), an infinity of {\em sideband modes} are
produced at the output, with frequencies shifted from $\omega $ by multiples
of $2(\omega -\omega _{s})$. We analyze the optical bistability of the two
nearest sideband modes, red-shifted and blue-shifted.
\end{abstract}


%

\section{Introduction}

%
Optical bistability (OB) has been the subject of intense research since its
prediction and observation in the 1970's \cite{szoke,mccall,mc2}. In Ref. 
\cite{boni1} a model consisting of a system of homogeneously broadened
two-level atoms driven by a coherent resonant field proved to give a
successful description of OB. Due its potential applications in optical
devices there has been a lot of efforts to observe and understand the
phenomenon of optical bistability in two-level atoms \cite
{abraham,milonni,haas1,galatola,bergou1,hassan12,hassan3,boni3,boni4,boni5,boni6,boni7,boni8}%
.

The effects of the squeezed vacuum field on the absorptive OB for a system
of two-level atoms in a ring cavity (see Figure 1), with different
relaxation rates of the in-quadrature and in-phase components, were
originally calculated in \cite{haas1}. The authors verified that the
squeezed vacuum strongly affects the OB, through the increase of the atomic
decay time and through the introduction of a relative phase between the
input pumping and squeezed vacuum fields. Although several aspects of
squeezed vacuum effects on OB have been considered \cite
{galatola,bergou1,hassan12,hassan3}, no explicit calculations where done, to
our knowledge, to the situation where the frequencies of the input fields,
pump ($\omega $) and broadband squeezed vacuum (carrier $\omega _{s}$) are
detuned. In papers \cite{galatola,bergou1,hassan12,hassan3}, exact resonance
between pump and squeezed fields frequencies, $\omega =\omega _{s}$, were
assumed in order to maximize the squeezing effects. Nonetheless,
consideration of detuning, $\omega \neq \omega _{s} $, is the source of
interesting physics as to be shown in this paper. Here, we analyze the
effects of that detuning over the OB in the output field, produced by a
system of two-level atoms in a cavity. We show that, besides the central
mode at $\omega $, the output field contains an infinity of sideband modes
at frequencies shifted from $\omega $ by multiples of $2(\omega -\omega
_{s}) $. We analyze the OB of the two nearest sideband modes, red-shifted
and blue-shifted.

The paper is organized as follows: In Sec. II we introduce the model we use,
and derive the system dynamical equations. In Sec. III we obtain the
stationary solutions for the output field. In Sec. IV we discuss the results
and present our conclusions. Finally, in Appendix A we derive the many-body
master equation and apply the mean-field approximation for a dilute atomic
gas. %

\section{Model}

%
We consider an input pump coherent signal of undepleted electric field
amplitude $E_{in}$ and a broadband squeezed vacuum, with frequency
distribution centered at $\omega _{s}$, interacting with $N$ two-level
atoms. The Hamiltonian of the system is given by 
\begin{equation}
H=\frac{1}{2}\omega _{0}S_{0}+F^{\ast }e^{i\omega t}S_{-}+Fe^{-i\omega
t}S_{+}+\sum_{k }\omega _{k}b_{k}^{+}b_{k}+\sum_{k}\left(
g_{k}b_{k}S_{+}+g_{k}^{\ast }b_{k}^{+}S_{-}\right) ,  \label{ham1}
\end{equation}
(we have set $\hbar =1$), where, the first term stands for the two-level
atomic system (transition frequency $\omega _{0}$), the two following terms
represent the interaction between the atoms and the input pump field
amplitude $E_{in}$, $F=\mu E_{in}$ ( $\mu $ is the atomic dipole moment),
the fourth term corresponds to the squeezed vacuum modes and the last one is
for the interaction between atoms and squeezed vacuum field. Operator $b_{k}$
($b_{k}^{+}$) annihilates (creates) squeezed field quanta of frequency $%
\omega _{k}$ and $g_{k}$ is the coupling constant. The atomic collective
operators are 
\begin{equation}
S_{0}=\sum_{i=1}^{N}s_{0}(i);\quad S_{\pm }=\sum_{i=1}^{N}s_{\pm }(i),
\label{opercol}
\end{equation}
where $s_{0}(i)$ and $s_{\pm }(i)$ are single particle operators satisfying
the commutation relations $\left[ s_{0}(i),s_{\pm }(j)\right] =\pm 2\delta
_{i,j}s_{\pm }(i)$ and $\left[ s_{+}(i),s_{-}(j)\right] =\delta
_{i,j}s_{0}(i)$. Although the atoms do not interact directly with each other
and the coherent field is assumed undepleted, they become correlated to each
other, only due to their coupling with the squeezed vacuum field.

In the mean field approximation and in a rotating frame at frequency $\omega 
$, the atomic system is described by an one-body master equation, obtained
by calculating the trace over the squeezed vacuum degrees of freedom (see
Appendix A for a detailed derivation), 
\begin{eqnarray}
\frac{d\rho (t)}{dt} &=&\frac{1}{i}\left[ H_{eff},\rho (t)\right] -\left\{ %
\left[ e^{i\theta }e^{i\epsilon t}\left( \frac{\gamma }{2}-i\nu \right)
2\sinh r\cosh r\ s_{+}\rho s_{+}+h.c.\right] \right.  \nonumber \\
&&\left. +\frac{\gamma }{2}\sinh ^{2}r\left( s_{-}s_{+}\rho -2s_{+}\rho
s_{-}+\rho s_{-}s_{+}\right) +\frac{\gamma }{2}\cosh ^{2}r\left(
s_{+}s_{-}\rho -2s_{-}\rho s_{+}+\rho s_{+}s_{-}\right) \right\} .
\label{master1}
\end{eqnarray}
\noindent The term in braces represents the phase-sensitive damping due the
squeezed vacuum, $r$ is the squeezing parameter, $\epsilon =2\left( \omega
-\omega _{s}\right) $ is twice the detuning between input pump and squeezed
vacuum fields, $\gamma $ is the damping constant and $\theta $ is a phase
reference of the squeezed vacuum field. $H_{eff}$ is an effective nonlinear
mean-field single particle Hamiltonian, describing the motion of one atom in
the sample, 
\begin{equation}
H_{eff}=\frac{1}{2}\left( \delta -\nu \cosh 2r\right) s_{0}+\mu \left\{ %
\left[ E_{in}+\frac{N-1}{\mu }\left( -\nu +i\frac{\gamma }{2}\right)
\left\langle s_{+}\right\rangle \right] s_{-}+h.c.\right\} ,  \label{hamil1}
\end{equation}
where $\left\langle s_{\pm }\right\rangle ={\rm Tr}(\rho s_{\pm })$, $\delta
=\omega _{0}-\omega $ is the detuning between atomic transition and pump
field frequencies and $\nu $ is the dynamic frequency shift; being much
smaller than $\gamma $, it will be neglected \cite{gardiner}. From second
term in the hamiltonian (\ref{hamil1}) we see that effectively a single
generic atom is excited by the imput field $E_{in}$ plus a polarization
field 
\begin{equation}
\epsilon _{pol}(t)=\frac{N-1}{\mu }\left( -\nu +i\frac{\gamma }{2}\right)
\left\langle s_{+}\right\rangle
\end{equation}
due the other $(N-1)$ atoms.

The equations of motion for the atomic operators mean values are 
\begin{equation}
\langle \dot{s_{0}}\rangle =2i\mu \left( {\epsilon }_{T}(t)\langle
s_{-}\rangle -{\epsilon }_{T}^{\ast }(t)\langle s_{-}\rangle ^{\ast }\right)
-\gamma \left( \langle s_{0}\rangle \cosh 2r+1\right) ,  \label{s0}
\end{equation}
\begin{equation}
\langle \dot{s_{-}}\rangle =-i\Omega \langle s_{-}\rangle +i\mu {\epsilon }%
_{T}^{\ast }(t)\langle s_{0}\rangle -Qe^{i\epsilon t}\langle s_{-}\rangle
^{\ast },  \label{s-}
\end{equation}
where $\langle \dot{s_{+}}\rangle =\langle \dot{s_{-}}\rangle ^{\ast }$, $%
\Omega \equiv \delta -i(\gamma /2)\cosh 2r$ , $Q\equiv (\gamma /2)e^{i\theta
}\sinh 2r$, and 
\begin{equation}
\epsilon _{T}(t)=E_{in}+\epsilon _{pol}(t)  \label{campototal}
\end{equation}
is the total effective field experienced by a single atom. The second term
in $\Omega $ is due to the commutation relations in the Heisenberg
equations. Furthermore, in the induced atomic polarization field 
\begin{equation}
\epsilon _{pol}(t)\equiv \Lambda \langle s_{-}\rangle ^{\ast }(t)/\mu
,\qquad \left( \Lambda =i\frac{\gamma }{2}N_{eff}\right)
\label{campointerno}
\end{equation}
we have assumed an effective number of atoms $N_{eff}$ contributing
effectively to this field ($N_{eff}\ll N$).

In the next section we obtain the stationary effective field amplitude $%
\epsilon _{T}(t)$ as function of $E_{in}$ and system parameters. %

\section{Stationary solutions}

%
For no detuning between vacuum squeezed and pump fields, $\epsilon =0$,
there is no explicit time dependence in Eq. (\ref{s-}), and the equilibrium
solutions ($\langle \dot{s_{-}}\rangle =0$, $\langle \dot{s_{0}}\rangle =0$) 
$\langle s_{-}\rangle ^{eq}$ and $\langle s_{0}\rangle ^{eq}$ are easily
obtained as function of the output field $\epsilon _{T}$, which, together
with Eq. (\ref{campototal}) enables to recover the well known result \cite
{haas1,bergou1}, 
\begin{equation}
E_{in}={\epsilon }_{T}-\frac{\gamma \mu \Lambda \left( \Omega {\epsilon }%
_{T}-iQ^{\ast }{\epsilon }_{T}^{\ast }\right) }{4 \mu ^2\Omega _{I}\left| {%
\epsilon }_{T}\right| ^{2}-2 \mu^2 \left( Q{\epsilon }_{T}^{2}+Q^{\ast
}\left( {\epsilon } _{T}^{\ast }\right) ^{2}\right) -\gamma \left( \left|
\Omega \right| ^{2}-\left| Q\right| ^{2}\right) \cosh 2r}.
\end{equation}
The bistable behavior becomes evident from plotted output field amplitude
modulus $\left| \epsilon _{T}\right| $ as function of the same for the pump
field $\left|E_{in}\right|$, as displayed in Figure 2. Above a critical
value of $N_{eff}$ an $S$-shaped curve is produced, meaning that there are
two possible output fields for a single input one. Moreover, the $S$-shaped
curve is quite sensible to the phase difference between input and squeezed
vacuum fields, as stressed in Refs. \cite{haas1,bergou1}.

We are interested in the situation $\epsilon \neq 0$, when equations (\ref
{s0})-(\ref{s-}) are no more autonomous, so the asymptotic stationary
solutions are periodical time-dependent series 
\begin{equation}
\langle s_{-}\rangle =\sum_{n=-\infty }^{\infty }a_{n}e^{in\epsilon t},\quad
\langle s_{0}\rangle =\sum_{n=-\infty }^{\infty }b_{n}e^{in\epsilon t},
\label{expansao}
\end{equation}
whose coefficients can be determined from equation (\ref{s0}) and (\ref{s-}%
). The output field amplitude is also expanded as an infinite series 
\begin{equation}
{\epsilon }_{T}(t)=\sum_{n=-\infty }^{\infty }{\cal E}_{n}e^{in\epsilon t}.
\label{expansao2}
\end{equation}
In a non-rotating frame, the total output field amplitude is a superposition
of an infinite and countable number of modes, 
\begin{equation}
E_{T}(t)=\epsilon _{T}(t)e^{-i\omega t}={\cal E}_{0}e^{-i\omega t}+{\cal E}
_{+1}e^{-i(\omega -\epsilon )t}+{\cal E}_{-1}e^{-i(\omega +\epsilon
)t}+...\, ,
\end{equation}
at frequencies $\omega _{n}=\omega \pm n\epsilon $, for $n=0,1,2,...$.

Inserting the series (\ref{expansao}) and (\ref{expansao2}) into Eqs. (\ref
{s0})-(\ref{s-}) and equalling coefficients with same time dependent factor $%
e^{in\epsilon t}$, one gets the following equations in terms $a_{n}$ and $%
b_{n}$, 
\begin{equation}
E_{in}={\cal E}_{0}-\Lambda a_{0}^{\ast },\qquad {\rm {for\quad }}n=0,
\label{ef0}
\end{equation}
\begin{equation}
{\cal E}_{n}=\Lambda a_{-n}^{\ast },\quad {\rm {for\quad }}n\neq 0.
\label{efn}
\end{equation}
\begin{equation}
i\left( n\epsilon +\Omega \right) a_{n}+Qa_{-n+1}^{\ast }=i\sum_{m=-\infty
}^{\infty }{\cal E}_{m-n}^{\ast }b_{m},\qquad ,  \label{coef1}
\end{equation}
\begin{eqnarray}
&&(in\epsilon +\gamma \cosh 2r)b_{n}  \nonumber \\
&=&2i\sum_{m=-\infty }^{\infty }\left( {\cal E}_{n-m}a_{m}-{\cal E}%
_{m-n}^{\ast }a_{-m}^{\ast }\right) -\gamma \delta _{n,0}.  \label{coef2}
\end{eqnarray}
After a lengthy but straightforward algebraic manipulation of equations (\ref
{efn})-(\ref{coef2}), one obtains an equation involving only the
coefficients $a_{n}$ and the central output field amplitude ${\cal E}_{0}$, 
\begin{eqnarray}
&&G_{n}(\epsilon )a_{n}+F_{n}(\epsilon )a_{-n}^{\ast }+Qa_{-n+1}^{\ast }+i%
\frac{{\cal E}_{o}^{\ast }}{\cosh 2r}\delta _{n,0}=-2\sum_{l(\neq n)}\left\{ 
\frac{{\cal E}_{o}^{\ast }}{Y_{n}(\epsilon )}\left[ \Lambda a_{l-n}^{\ast
}a_{l}-\Lambda ^{\ast }a_{n-l}a_{-l}^{\ast }\right] \right.   \nonumber \\
&+&\left. \frac{\Lambda ^{\ast }}{Y_{l}(\epsilon )}\left[ {\cal E}%
_{o}a_{n-l}a_{n}-{\cal E}_{o}^{\ast }a_{n-l}a_{-n}^{\ast }\right]
+\sum_{m(\neq n)}\frac{1}{Y_{m}(\epsilon )}\left[ \left| \Lambda \right|
^{2}a_{n-m}a_{l-n}a_{l}-(\Lambda ^{\ast })^{2}a_{n-m}a_{n-l}a_{-l}^{\ast }%
\right] \right\}   \label{nonlinear}
\end{eqnarray}
where 
\begin{eqnarray}
&&G_{n}(\epsilon )=i\left[ n\epsilon +\Omega +\frac{\gamma \Lambda ^{\ast
}(1-\delta _{n,0})}{Y_{0}(\epsilon )}\right] +2\frac{\left| {\cal E}%
_{o}\right| ^{2}}{Y_{n}(\epsilon )},  \label{G1} \\
&&F_{n}(\epsilon )=-2\frac{({\cal E}_{0}^{\ast })^{2}}{Y_{n}(\epsilon )}
\label{F1}
\end{eqnarray}
and
\begin{equation}
Y_{n}(\epsilon )=in\epsilon +\gamma \cosh 2r.  \label{Y1}
\end{equation}
On the left-hand-side (LHS) of Eq. (\ref{nonlinear}) $N_{eff}$ enters only
in $G_{n}(\epsilon )$, while on right-hand-side (RHS) it enters the terms
involving the products of $a_{n}$'s. For field intensities of sideband modes
much weaker than the central mode, we neglect the nonlinear terms on the RHS
of Eq. (\ref{nonlinear}). This allows us to rewrite the LHS in terms of a
finite difference equation for $a_{n}$, 
\begin{eqnarray}
&&B_{n}(\epsilon )a_{n}+C_{n}(\epsilon )a_{n+1}+D_{n}(\epsilon )a_{n-1} 
\nonumber \\
&=&E_{0}(\epsilon )\delta _{n,0}+H_{1}(\epsilon )\delta _{n,1},
\label{linear}
\end{eqnarray}
where 
\begin{eqnarray}
B_{n}(\epsilon ) &=&G_{n}(\epsilon )-\frac{F_{n}(\epsilon )F_{-n}^{\ast
}(\epsilon )}{G_{-n}^{\ast }(\epsilon )}-\frac{|Q|^{2}}{G_{-n+1}^{\ast
}(\epsilon )},  \label{B} \\
C_{n}(\epsilon ) &=&-\frac{Q^{\ast }F_{n}(\epsilon )}{G_{-n}^{\ast
}(\epsilon )},  \label{C} \\
D_{n}(\epsilon ) &=&-\frac{QF_{-n+1}^{\ast }(\epsilon )}{G_{-n+1}^{\ast
}(\epsilon )},  \label{D} \\
E_{n}(\epsilon ) &=&-\frac{i}{\cosh 2r}\left[ \frac{{\cal E}%
_{o}F_{n}(\epsilon )}{G_{-n}^{\ast }(\epsilon )}+{\cal E}_{o}^{\ast }\right]
,  \label{E} \\
H_{n}(\epsilon ) &=&-i\frac{Q{\cal E}_{o}}{G_{-n+1}^{\ast }(\epsilon )\cosh
2r}.  \label{H}
\end{eqnarray}

Even in this very linear approximation the $n$-dependence in the
coefficients (\ref{B})-(\ref{H}) does not allow obtaining an exact closed
solution to Eq. (\ref{linear}), for $\epsilon \neq 0$. In the present
analysis, we are going to determine only the first three sidebands
coefficients $a_{0}$ and $a_{\pm 1}$. From Eq. (\ref{linear}) one gets the
following system of equations 
\begin{eqnarray}
B_{0}a_{0}+C_{0}a_{1}+D_{0}a_{-1} &=&E_{0}  \nonumber \\
B_{1}a_{1}+C_{1}a_{2}+D_{1}a_{0} &=&H_{1}  \label{system} \\
B_{-1}a_{-1}+C_{-1}a_{0}+D_{-1}a_{-2} &=&0,  \nonumber
\end{eqnarray}
which is not closed because $a_{0}$ and $a_{\pm 1}$ are coupled to $a_{\pm
2} $, that, by their turn, are coupled to higher order coefficients. Instead
of simply disregarding $a_{2}$ and $a_{-2}$ in Eqs. (\ref{system}), we
consider a better approximation by estimating them from truncated continued
fractions. Setting 
\begin{equation}
x_{n}\equiv \frac{a_{n}}{a_{n-1}},\quad y_{-n}\equiv \frac{a_{-(n+1)}}{a_{-n}%
}\quad \,,
\end{equation}
\noindent for $a_{n-1}\neq 0,a_{-n}\neq 0$ and $n\neq 0,1$ we can write Eq. (%
\ref{linear}) as two equations, 
\begin{equation}
x_{n}=\frac{-D_{n}}{B_{n}+C_{n}x_{n+1}},\quad n\neq 0,1,  \label{fracx}
\end{equation}
\begin{equation}
y_{-n}=\frac{-C_{-(n+1)}}{B_{-(n+1)}+D_{-(n+1)}y_{-(n+1)}},  \label{fracy}
\end{equation}
for positive integers $n$. For $n=2$ in (\ref{fracx}), $n=1$ in (\ref{fracy}%
) and truncation of the continued fractions, up to a second order iteration,
yields (a higher order iteration does not affect significantly the result) 
\begin{equation}
\frac{a_{2}}{a_{1}}=x_{2}^{(2)}=-\frac{D_{2}}{B_{2}-\frac{C_{2}D_{3}}{B_{3}-%
\frac{C_{3}D_{4}}{B_{4}}}}  \label{fra1a}
\end{equation}
\begin{equation}
\frac{a_{-2}}{a_{-1}}=y_{-1}^{(2)}=-\frac{C_{-2}}{B_{-2}-\frac{D_{-2}C_{-3}}{%
B_{-3}-\frac{D_{-3}C_{-4}}{B_{-4}}}}.  \label{fra2b}
\end{equation}
Substituting $a_{2}=x_{2}^{(2)}a_{1}$, $a_{-2}=y_{-1}^{(2)}a_{-1}$, we get
the coefficients 
\begin{equation}
a_{0}=\frac{E_{0}-\frac{C_{0}H_{1}}{B_{1}+C_{1}x_{2}^{(2)}}}{B_{0}-\frac{%
D_{0}C_{-1}}{B_{-1}+D_{-1}y_{-1}^{(2)}}-\frac{C_{0}D_{1}}{%
B_{1}+C_{1}x_{2}^{(2)}}},  \label{a0}
\end{equation}
\begin{equation}
a_{1}=\frac{H_{1}-D_{1}\,a_{0}}{B_{1}+C_{1}x_{2}^{(2)}},\quad a_{-1}=-\frac{%
C_{-1}\,a_{0}}{B_{-1}+D_{-1}y_{-1}^{(2)}}.  \label{a1}
\end{equation}
%

\section{Results and conclusions}

%
Using the solutions for the amplitudes $a_{0}$ and $a_{\pm 1}$, Eqs. (\ref
{a0}) and (\ref{a1}), we can analyze the functional dependence of the output
fields ${\cal E}_{0}$ and ${\cal E}_{\pm }$ as function of the input field $%
E_{in}$, in modulus. To simplify the illustration of the bistable behavior,
we assume $\theta $ being the phase difference between pump and squeezed
input fields. The output field amplitudes $\left| {\cal E}_{0}\right| $, $%
\left| {\cal E}_{+1}\right| =\left| \Lambda a_{-1}^{\ast }\right| $ and $%
\left| {\cal E}_{-1}\right| =\left| \Lambda a_{1}^{\ast }\right| $ are
plotted as functions of $E_{in}$ in Figs. 3-(a),3-(b), and 3-(c)
respectively. The parameters are set as $N_{eff}=101$, $\epsilon /\gamma
=2.0 $, $r=0.5$, $\delta =0$, and $\theta =\pi $. We verified that the OB
looses the phase-sensitivity, varying more significantly with $r$, because
the coefficients $a_{0}$, $a_{1}$ and $a_{-1}$ now depend on $|Q|^{2}$,
instead on $Q$. Although the sideband field intensities are much weaker than
the central one, they also display a bistable behavior, with turning points
occurring at the same input field intensity. The dashed lines correspond to
the unstable branches, the arrows indicate the path followed by the output
field variation as the input is increased or decreased. The bistable
behavior of the central mode (Fig. 3-(a)) is similar the case where $%
\epsilon =0$, however, the sideband modes, ${\cal E}_{1}$ (Fig. 3-(b)) and $%
{\cal E}_{-1}$ (Fig. 3-(c)), that are respectively, red-shifted and
blue-shifted with respect to the central mode, show some qualitative
differences. Differently from the central mode, at strong pump amplitude
modulus, the sideband modes show a monotonic decrease in the amplitude
modulus at the output. The sideband modes also present the following
different features in the switchings, or jumps from low to high amplitude
(and vice-versa) in comparison with the central mode: i) By increasing the
input field intensity the $(a)\rightarrow (b)$ switch is from low to high
amplitude, in modes ${\cal E}_{0}$ and ${\cal E}_{+1}$, see Figs. 3-(a) and
3-(b), however it is inverted in mode ${\cal E}_{-1}$, switching from high
to low amplitude, see Fig. 3-(c). ii) By reverting the path, going from high
to low input intensity the switches occur from high to low output
amplitudes, $(c)\rightarrow (d)$, in modes ${\cal E}_{0}$ and ${\cal E}_{+1}$%
, Figs. 3-(a) and 3-(b), while it is from low to high in mode ${\cal E}_{-1}$%
. Essentially, the sideband modes show inverse behavior, with respect to the
switchings. iii) Comparatively to the central mode, the sidebands present a
higher contrast in the jumps from higher to lower amplitude.

A possible application of the above results could be the simultaneous
transmission of a message by the output field through three different
channels (the three modes), where the triplicated information could be
useful for error control. Additionally, the codification in the blue-shifted
sideband (0,1,1,0,0,1,...) is the inverse of that in the other mode
(1,0,0,1,1,0,...), so the sideband modes could transmit information as like
the codification occurring in the DNA double-strand macromolecule, where one
strand sequence is the inverse of the other.

In conclusion, we have shown that the frequency detuning between input pump
and squeezed vacuum fields, interacting with two-level atoms, gives rise to
a multiple-mode output field with frequencies that are multiples of $%
\epsilon $. By analyzing the closest (red-shifted and blue-shifted) sideband
modes, to the central one, we did verify new features in the bistable
behavior. Although the obtention of a pump and squeezed fields with
controllable phase difference could be, at the moment, experimentally
difficult, because both should derive from a common source, we believe that
the reported physical effects could be useful in optical devices and in the
transmission of information. %
\acknowledgments{LPM, GAP and SSM, acknowledge financial support from
FAPESP (S\~ao Paulo, SP, Brazil), under contracts \# 00/15084-5,
99/11129-5. SSM also acknowledges partial financial support from CNPq
(DF, Brazil).} 
%
\appendix
%

\section{Master equation}

%
In a referential frame rotating at frequencies $\omega -\omega _{k}$ the
hamiltonian (\ref{ham1}) becomes 
\begin{equation}
H=H_{0S}+V(t),  \label{ham2}
\end{equation}
where 
\begin{equation}
H_{0S}=\frac{\delta }{2}S_{0}+F^{\ast }S_{-}+FS_{+},  \label{ham3}
\end{equation}
the collective operators are defined in (\ref{opercol}), $\delta = \omega -
\omega_0$ and 
\begin{equation}
V(t)=\sum_{k }\left( g_{k }b_{k}S_{+}e^{i\left( \omega -\omega _{k}\right)
t}+h.c.\right) .  \label{int1}
\end{equation}

Following the usual procedure \cite{gardiner}, by eliminating the reservoir
degrees of freedom one obtains a pre-master equation for the system density
operator $\rho (t)$ 
\begin{equation}
\frac{d\rho (t)}{dt}=\frac{1}{i}\left[ H_{0S},\rho (t)\right]
-\int_{0}^{t}dt^{\prime }{\rm Tr}_{{\cal R}}\left[ V(t),\left[ V(t^{\prime
}),\rho (t^{\prime })\rho _{R}\right] \right]  \label{eqmaster1}
\end{equation}
where $\rho _{R}$ is the state of the reservoir, at thermal equilibrium.
Substituting the interaction (\ref{int1}) in (\ref{eqmaster1}) one gets 
\begin{eqnarray}
\frac{d\rho (t)}{dt} &=&-i\left[ H_{0S},\rho (t)\right] -\int_{0}^{t}dt^{%
\prime }\left\{ \xi _{11}(t,t^{\prime })\left[ S_{+},\left[ S_{+},\rho
(t^{\prime })\right] \right] +\xi _{12}^{\ast }(t,t^{\prime })\left[
S_{-},S_{+}\rho (t^{\prime })\right] \right.  \label{eqmaster2} \\
&&\left. +\xi _{21}(t,t^{\prime })\left[ S_{+},S_{-}\rho (t^{\prime })\right]
+h.c.\right\} .
\end{eqnarray}
The coefficients $\xi _{ij}(t,t^{\prime })$ are characterized by the kind of
reservoir, 
\begin{equation}
\int_{0}^{t}dt^{\prime }\xi _{11}(t,t^{\prime })\rho (t^{\prime
})=\int_{0}^{t}dt^{\prime }\sum_{\omega ,^{\prime }\omega ^{\prime \prime
}}g_{\omega ^{\prime }}g_{\omega ^{\prime \prime }}e^{i\left( \omega -\omega
^{\prime }\right) t+i\left( \omega -\omega ^{\prime \prime }\right)
t^{\prime }}\left\langle b_{\omega ^{\prime }}b_{\omega ^{\prime \prime
}}\right\rangle _{R}\rho (t^{\prime }),  \label{cor1}
\end{equation}
\begin{equation}
\int_{0}^{t}dt^{\prime }\xi _{12}(t,t^{\prime })\rho (t^{\prime
})=\int_{0}^{t}dt^{\prime }\sum_{\omega ,^{\prime }\omega ^{\prime \prime
}}g_{\omega ^{\prime }}^{\ast }g_{\omega ^{\prime \prime }}e^{-i\left(
\omega -\omega ^{\prime }\right) t+i\left( \omega -\omega ^{\prime \prime
}\right) t^{\prime }}\left\langle b_{\omega ^{\prime }}^{+}b_{\omega
^{\prime \prime }}\right\rangle _{R}\rho (t^{\prime }),  \label{cor2}
\end{equation}
\begin{equation}
\int_{0}^{t}dt^{\prime }\xi _{21}(t,t^{\prime })\rho (t^{\prime
})=\int_{0}^{t}dt^{\prime }\sum_{\omega ,^{\prime }\omega ^{\prime \prime
}}g_{\omega ^{\prime }}g_{\omega ^{\prime \prime }}^{\ast }e^{i\left( \omega
-\omega ^{\prime }\right) t-i\left( \omega -\omega ^{\prime \prime }\right)
t^{\prime }}\left\langle b_{\omega ^{\prime }}b_{\omega ^{\prime \prime
}}^{+}\right\rangle _{R}\rho (t^{\prime }).  \label{cor3}
\end{equation}
where $\left\langle b_{\omega ^{\prime }}b_{\omega ^{\prime \prime
}}\right\rangle _{R}={\rm Tr}_{R}\left( \rho _{R}b_{\omega ^{\prime
}}b_{\omega ^{\prime \prime }}\right) $. For a squeezed reservoir 
\begin{equation}
\left\langle b_{\omega ^{\prime }}b_{\omega ^{\prime \prime }}\right\rangle
_{R}=-e^{i\theta }\sinh r\cosh r\ \delta \left[ \omega ^{\prime \prime
}-\left( 2\omega _{s}-\omega ^{\prime }\right) \right] ;\qquad  \label{cor4}
\end{equation}
\begin{equation}
\left\langle b_{\omega ^{\prime }}^{+}b_{\omega ^{\prime \prime
}}\right\rangle _{R}=\sinh ^{2}r\ \delta \left( \omega ^{\prime }-\omega
^{\prime \prime }\right) ;\ \ \left\langle b_{\omega ^{\prime }}b_{\omega
^{\prime \prime }}^{+}\right\rangle _{R}=\cosh ^{2}r\ \delta \left( \omega
^{\prime }-\omega ^{\prime \prime }\right) \   \label{cor5}
\end{equation}
where $r$ is the squeeze parameter, $\theta $ is the reference phase of the
squeezed field, and $\omega _{s}$ is the central resonant frequency of the
squeezing device. Going from sums to integrals in Eqs. (\ref{cor1}-\ref{cor3}%
) and using expressions (\ref{cor4}-\ref{cor5}), one gets for example 
\[
\int_{0}^{t}dt^{\prime }\xi _{11}(t,t^{\prime })\rho (t^{\prime
})=-e^{i\theta }\sinh r\cosh r\int_{0}^{\infty }d\omega ^{\prime }D(\omega
^{\prime })g(\omega ^{\prime })g(2\omega _{s}-\omega ^{\prime })e^{i(\omega
-\omega ^{\prime })t}\int_{0}^{t}dt^{\prime }e^{i(\omega -2\omega
_{s}+\omega ^{\prime })t^{\prime }}\rho (t^{\prime }) 
\]
where $D(\omega )$ is the reservoir density of modes. Making the change $%
t-t^{\prime }=$ $\tau $ and invoking the Markov approximation $\rho (t-\tau
)\simeq \rho (t)$ we obtain 
\[
\int_{0}^{t}d\tau e^{-i(\omega -2\omega _{s}+\omega ^{\prime })\tau }\rho
(t-\tau )\cong \int_{0}^{\infty }d\tau e^{-i(\omega -2\omega _{s}+\omega
^{\prime })\tau }\rho (t)=\rho (t)\left[ \pi \delta \left( \omega -2\omega
_{s}+\omega ^{\prime }\right) -i{\cal P}\frac{1}{\omega -2\omega _{s}+\omega
^{\prime }}\right] , 
\]
and 
\[
\int_{0}^{t}dt^{\prime }\xi _{11}(t,t^{\prime })\rho (t^{\prime })\cong 
\tilde{\xi}_{11}(t)\rho (t) 
\]
with 
\[
\tilde{\xi}_{11}(t)=-e^{i\theta }\sinh r\cosh r\ e^{2i(\omega -\omega _{s})t}%
\left[ \pi D(2\omega _{s}-\omega )g(2\omega _{s}-\omega )g(\omega )-i{\cal P}%
\int_{0}^{\infty }d\omega ^{\prime }\frac{D(\omega ^{\prime })g(\omega
^{\prime })g(2\omega _{s}-\omega ^{\prime })}{\omega -2\omega _{s}+\omega
^{\prime }}\right] 
\]
where ${\cal P}$ stands for the Cauchy principal value. For $\left| \omega
_{s}-\omega \right| \ll \omega $ the two terms in the brackets are assumed
being approximately constant, so we define the damping constant ($\gamma $)
and the dynamical frequency shift ($\nu _{s}$) 
\[
\gamma \equiv 2\pi Dg^{2},\qquad \nu \equiv {\cal P}\int_{0}^{\infty
}d\omega ^{\prime }\frac{D(\omega ^{\prime })g(\omega ^{\prime })g(2\omega
_{s}-\omega ^{\prime })}{\omega -2\omega _{s}+\omega ^{\prime }}, 
\]
therefore 
\[
\tilde{\xi}_{11}(t)=-e^{i\theta }\sinh r\cosh r\ e^{2i(\omega -\omega
_{s})t}\left( \frac{\gamma }{2}-i\nu \right) . 
\]
Following the same procedure one obtains the other coefficients, 
\[
\tilde{\xi}_{12}=\left( \frac{\gamma }{2}-i\nu _{s}\right) \sinh
^{2}r;\qquad \tilde{\xi}_{21}=\left( \frac{\gamma }{2}-i\nu \right) \cosh
^{2}r, 
\]
which are time-independent.

Thus the master equation for an $N$-atom system becomes 
\begin{equation}
\frac{d\rho _{N}(t)}{dt}=\frac{1}{i}\left[ H_{0S}^{(N)},\rho _{N}(t)\right]
-\left\{ \tilde{\xi}_{11}(t)\left[ S_{+},\left[ S_{+},\rho _{N}(t)\right] %
\right] +\tilde{\xi}_{12}^{\ast }\left[ S_{-},S_{+}\rho _{N}(t)\right] +%
\tilde{\xi}_{21}\left[ S_{+},S_{-}\rho _{N}(t)\right] +h.c.\right\} ,
\label{eqmaster3}
\end{equation}
while for a system of $p-$atom system, $p<N$, it is 
\[
\frac{d\rho _{p}(t)}{dt}=-i\left[ H_{0S}^{(p)},\rho _{p}(t)\right] -\left\{ 
\tilde{\xi}_{11}(t)\sum_{i,j=1}^{p}\left[ s_{+}(i),\left[ s_{+}(j),\rho
_{p}(t)\right] \right] \right. 
\]
\[
+\tilde{\xi}_{12}^{\ast }\left( \sum_{i,j=1}^{p}\left[ s_{-}(i),s_{+}(j)\rho
_{p}(t)\right] +\left( N-p\right) \sum_{i=1}^{p}\left[ s_{-}(i),{\rm Tr}%
_{p+1}s_{+}(p+1)\rho _{p+1}(t)\right] \right) 
\]
\begin{equation}
+\left. \tilde{\xi}_{21}\left( \sum_{i,j=1}^{p}\left[ s_{+}(i),s_{-}(j)\rho
_{p}(t)\right] +\left( N-p\right) \sum_{i=1}^{p}\left[ s_{+}(i),{\rm Tr}%
_{p+1}{\rm \ }s_{-}(p+1)\rho _{p+1}(t)\right] \right) +h.c.\right\} .
\label{eqmaster4}
\end{equation}
For a dilute system the atomic correlations may be disregarded, so, we shall
consider a single generic atom ($p=1)$ moving in a mean field produced by
all the others, with the $2-$atom density operator factorized as $\rho
_{2}\approx \rho _{1}\otimes \rho _{1}$. In this approximation equation (\ref
{eqmaster4}) reduces to 
\[
\frac{d\rho _{1}(t)}{dt}=\frac{1}{i}\left[ H_{0S}^{(1)},\rho _{1}(t)\right]
-\left\{ \tilde{\xi}_{11}(t)\left[ s_{+},\left[ s_{+},\rho _{1}(t)\right] %
\right] +\tilde{\xi}_{12}^{\ast }\left( \left[ s_{-},s_{+}\rho _{1}(t)\right]
+\left( N-1\right) \left\langle s_{+}\right\rangle \left[ s_{-},\rho _{1}(t)%
\right] \right) \right. 
\]
\begin{equation}
+\left. \tilde{\xi}_{21}\left( \left[ s_{+},s_{-}\rho _{1}(t)\right] +\left(
N-1\right) \left\langle s_{-}\right\rangle \left[ s_{+},\rho _{1}(t)\right]
\right) +h.c.\right\} .  \label{eqmaster5}
\end{equation}
with the single particle Hamiltonian 
\begin{equation}
H_{0S}^{(1)}=\frac{\delta }{2}s_{0}+F^{\ast }s_{-}+Fs_{+}  \label{ham5}
\end{equation}
and $\left\langle s_{\pm }\right\rangle ={\rm Tr}$($s_{\pm }\rho _{1})$ is
the mean value. Rearranging the terms in Eq. (\ref{eqmaster5}) and dropping
the subscript 1 in $\rho _{1}$ we can write Eq. (\ref{eqmaster5}) as 
\[
\frac{d\rho (t)}{dt}=-i\left[ H_{eff},\rho (t)\right] -\left\{ 2\left[
e^{i\theta }e^{2i\left( \omega -\omega _{s}\right) t}\left( \frac{\gamma }{2}%
-i\nu \right) \sinh r\cosh r\ s_{+}\rho s_{+}+h.c.\right] \right. 
\]
\begin{equation}
\left. +\frac{\gamma }{2}\sinh ^{2}r\left( s_{-}s_{+}\rho -2s_{+}\rho
s_{-}+\rho s_{-}s_{+}\right) +\frac{\gamma }{2}\cosh ^{2}r\left(
s_{+}s_{-}\rho -2s_{-}\rho s_{+}+\rho s_{+}s_{-}\right) \right\} .
\label{eqmaster6}
\end{equation}
The single particle effective Hamiltonian in Eq. (\ref{eqmaster6}) is given
by 
\begin{equation}
H_{eff}=\frac{1}{2}\left( \delta -\nu \cosh 2r\right) s_{0}+\left( F^{\ast
}s_{-}+Fs_{+}\right) +\left( N-1\right) \left[ \left( -\nu +i\frac{\gamma }{2%
}\right) \left\langle s_{+}\right\rangle s_{-}+h.c.\right] ,  \label{ham6}
\end{equation}
it contains nonlinear terms due the mean-field effect of the remaining $N-1$
atoms. The frequency shift $\nu \cosh 2r$ is due to the interaction with the
reservoir. The second term in the RHS of Eq. (\ref{eqmaster6}) stands for
the dissipative part due to the decay in the squeezed vacuum \cite{gardiner}%
. By setting $\omega =\omega _{s}$, identifying $\sinh r\cosh
r\longrightarrow \bar{m}$, $\sinh ^{2}r\longrightarrow \bar{n}$ and $\cosh
^{2}r\longrightarrow \bar{n}+1$, the dissipative term of the master equation
takes the same form as considered in \cite{haas1}.

%
\newpage

\begin{center}
{\Large FIGURE CAPTIONS}
\end{center}

\vspace{-2mm}

\noindent{\bf Figure 1.} Schematic ring cavity with N two-level atoms in a
cell. Input signal at the left, output at the right and injection of
squeezed vacuum from above. $M_1$ to $M_4$ specify the mirrors. \newline

\noindent{\bf Figure 2.} Output versus input field amplitudes. Sensitivity
to the phase $\theta$ is manifest. \newline

\noindent {\bf Figure 3.} Output modes versus input field $E_{in}$. Dashed
lines are for the unstable branches. Arrows indicate the direction of
variation of output amplitudes with increasing (decreasing) $E_{in}$. (a)
central mode amplitude ${\cal E}_{0}$. The jumps goes from $(a)\rightarrow
(b)$ ( $(c)\rightarrow (d)$ ) increasing (decreasing) the output amplitude.
(b) red-shifted sideband mode ${\cal E}_{+1}$. The jumps are in same
direction as in Figure (a). (c) blue-shifted sideband mode ${\cal E}_{-1}$.
The jump goes from $(a)\rightarrow (b)$ ($(c)\rightarrow (d)$) with
decreasing (increasing) amplitude of the output field. Both jumps occur in
direction opposite to those in Figures (a) and (b).The parameters are set as 
$N_{eff}=101$, $\epsilon /\gamma =2.0$, $r=0.5$, $\delta =0$, and $\theta
=\pi $


\end{document}